\documentclass[article]{revtex4}
\usepackage{graphicx}

\begin{document}
\title{Femtosecond nonlinear ultrasonics in gold probed with ultrashort surface plasmons}


\author{Vasily V. Temnov$^{1,2,\ast}$, Christoph Klieber$^{1}$, Keith A.
Nelson$^{1}$, Tim Thomay$^3$, Vanessa Knittel$^3$, Alfred
Leitenstorfer$^3$, Denys Makarov$^{4,5}$, Manfred Albrecht$^4$ and
Rudolf Bratschitsch$^4$}

\affiliation{$^1$Department of Chemistry, Massachusetts Institute
of Technology, Cambridge, Massachusetts 02139, USA}
\affiliation{$^2$Institut des Mol\'ecules et Mat\'eriaux du Mans,
UMR CNRS 6283, Universit\'e du Maine, 72085 Le Mans cedex, France}

\affiliation{$^3$Department of Physics and Center for Applied
Photonics, University of Konstanz, D-78457 Konstanz, Germany}
\affiliation{$^4$Institute of Physics, Chemnitz University of
Technology, 09107 Chemnitz, Germany}
\affiliation{$^5$Institute
for Integrative Nanosciences, IFW Dresden, 01069 Dresden, Germany}

\date{\today}

\maketitle \textbf{Fundamental interactions induced by lattice
vibrations on ultrafast time scales have become increasingly
important for modern nanoscience and technology. Experimental
access to the physical properties of acoustic phonons in the
terahertz frequency range and over the entire Brillouin zone is
crucial for understanding electric and thermal transport in solids
and their compounds. Here we report on the generation and
nonlinear propagation of giant (1 percent) acoustic strain pulses
in hybrid gold/cobalt bilayer structures probed with ultrafast
surface plasmon interferometry. This new technique allows for
unambiguous characterization of arbitrary ultrafast acoustic
transients. The giant acoustic pulses experience substantial
nonlinear reshaping after a propagation distance of only 100~nm in
a crystalline gold layer. Excellent agreement with the Korteveg-de
Vries model points to future quantitative nonlinear femtosecond
terahertz-ultrasonics at the nano-scale in metals at room
temperature.}

Investigations of nanometer-wavelength, terahertz-frequency
acoustic phonons usually require sophisticated methods with high
spatial resolution such as X-rays \cite{Sokolowski01PRL87},
electrons \cite{Park05PRB72}, or far ultraviolet radiation
\cite{Siemens10NMater9}. In metals, measurements at visible or
infrared wavelengths intrinsically bear the potential of nanometer
spatial resolution due to the characteristic penetration depth
(skin-depth) of optical fields of approximately 10~nm.
Time-resolved optical experiments in metal and semiconductor thin
films \cite{Thomsen86PRB34} provide a detailed picture of the
generation of picosecond acoustic pulses \cite{Saito03PRB67}.
However, the detection capabilities in picosecond ultrasonics are
generally limited by the largely unknown photo-elastic
coefficients leading to complicated dynamics of strain-induced
surface reflectivity, rendering the characterization of acoustic
pulses extremely challenging \cite{Devos01PRL86,Mante10PRB10}.
Some of these difficulties have been circumvented by using a more
straightforward beam deflection technique \cite{Wright92PRL69} or
equivalent interferometric detection of transient surface
deformations \cite{vanCapel10PRB81}. Nevertheless, the separation
of geometrical and photo-elastic contributions to the optical
phase still represents a problem. An even more complicated
experimental geometry utilizing joint phase- and
polarization-sensitive reflectivity measurements at oblique light
incidence was theoretically proposed to solve this problem
\cite{Matsuda03RSI74}. Inspired by the performance of the first
acousto-plasmonic pump-probe experiments in Kretschmann geometry
in 1988 \cite{vanExter88PRL60} and the recent progress in lateral
nanofabrication techniques in hybrid multilayer structures
combined with quantitative nanoplasmonic measurements below the
optical skin depth \cite{Temnov10NPhoton4,Temnov12NPhoton6}, we
have developed femtosecond surface plasmon interferometry
\cite{SPInterferometry09OE}. The phase of surface plasmons
propagating along the metal-dielectric interface is {\it not
sensitive} to transient surface displacements, but is only
affected by the strain-induced photo-elastic contribution
\cite{Temnov12NPhoton6}. As demonstrated below, this fact allows
us to fully characterize ultrashort acoustic pulses in gold by
monitoring transient modulations of the surface plasmon wave
vector.

Beyond these important metrological issues, recent studies have
revealed the possibility to generate ultrashort acoustic pulses by
magnetostriction \cite{KorffSchmising08PRB78}, to manipulate the
magnetization in ferromagnetic semiconductors
\cite{Scherbakov10PRL105,Thevenard10PRB82} and metals
\cite{Kim12PRL109} by ultrashort strain pulses, and to use
acoustic pulses to generate terahertz radiation
\cite{Armstrong09Nphys5}. These findings render the field of
terahertz acoustics truly interdisciplinary. Pioneering
investigations of {\it nonlinear} acoustic dynamics in dielectric
crystals at low temperature \cite{Hao01PRB64,vanCapel10PRB81} have
revealed clear signatures of shock-front decay into a sequence of
acoustic solitons after a propagation distance of hundreds of
microns. Surprisingly, acoustic nonlinearities have never been
investigated in metals, although they are used as sources of
large-amplitude acoustic pulses for these experiments. In this
article we demonstrate optical generation of large-amplitude
acoustic strain pulses in a hybrid gold/cobalt bilayer structure
and we observe nonlinear acoustic propagation effects in gold at
the nano-scale.

\textbf{\textsf{Results}}

\textbf{Acousto-plasmonics in a hybrid gold-cobalt multilayer
structure.}  We perform {\it quantitative} measurements by
plasmonic pump-probe interferometry with femtosecond time
resolution \cite{SPInterferometry09OE}. Figure 1a shows the
experimental configuration of this technique adopted to study
acousto-plasmonic effects. In contrast to most previous
measurements performed with a single polycrystalline metal layer
on a dielectric substrate, we use a Au/Co/sapphire structure with
well-defined crystallographic orientation. A hybrid
acousto-plasmonic 120~nm gold/35~nm cobalt/sapphire multilayer
structure was manufactured by magnetron sputtering of a (111) gold
layer on top of a hcp-cobalt film deposited on a (0001) sapphire
substrate. The cobalt layer excited by an intense ultrashort laser
pulse through the sapphire substrate serves as an efficient and
ultrafast opto-acoustic transducer. Due to strong electron-phonon
relaxation on the time scale of 200-300~fs
\cite{Koopmans2010NMat9} and the short electron mean free path
$l_{\rm e}\sim 1$~nm \cite{Getzlaff93SSC87} in ferromagnetic
materials, hot electron diffusion is much less efficient than in
noble metals \cite{Tas94PRB49}. Therefore, the heat penetration
depth only slightly exceeds the skin depth for pump light
(typically  by $\sim 50\%$ \cite{Saito03PRB67}), see Fig.~1a. The
laser-heated cobalt layer thermally expands and generates an
ultrashort acoustic strain pulse propagating both into the gold
layer and into the sapphire substrate. Excellent acoustic
impedance matching between the three layers suppresses acoustic
reflection at both the gold-cobalt and cobalt-sapphire interfaces
(about $10\%$). As a consequence, the initial shape of a unipolar
acoustic pulse follows the spatial profile of deposited heat in
cobalt. The nonlinear dynamics of ultrashort acoustic pulses in
sapphire have been addressed in detail before
\cite{Hao01PRB64,vanCapel10PRB81}. Here, we focus on nonlinear
acousto-plasmonic effects in a metal.

The compressional acoustic pulse in gold $\eta(z,t)=(n_{\rm
i}(z,t)-n_{\rm i}^0)/n_{\rm i}^0$ creates a layer of higher ion
density $n_{\rm i}(z,t)>n_{\rm i}^0$, which moves at the sound
velocity $c_{\rm s}$=3.45~km/s in gold in the (111) direction.
Since the stationary charge separation between electrons and ions
in a metal occurs only within the Debye radius $r_{\rm Debye}\sim
10^{-3}$~nm, the spatial profile of electron (charge) density
$n_{\rm e}(z,t)$ exactly follows the ionic one: $n_{\rm
e}(z,t)=n_{\rm i}(z,t)$. At the probe photon energy of 1.55~eV of
the probe pulses the dielectric function $\varepsilon_{\rm
m}=\varepsilon^{'}+i\varepsilon^{''}=-24.8+1.5i$ in gold is
dominated by the free-carrier contribution with
$\varepsilon^{'}\simeq -\omega^2_{\rm p}/\omega^2\propto -n_{\rm
e}$. An ultrashort acoustic strain pulse creates a time-dependent
spatial profile of the dielectric function $\varepsilon^{'}(z,t)=
\varepsilon^{'}(1+\eta(z,t))$ inside the metal, which modulates
the surface plasmon wavevector $k_{\rm
sp}=k_0\sqrt{\varepsilon_{\rm m}/(1+\varepsilon_{\rm m})}$, when
the strain pulse arrives within the surface plasmon skin depth
$\delta_{\rm skin}=13$~nm at the gold-air interface:
\begin{equation}
\label{AcPlasmEquation} \delta k_{\rm
sp}(t)=-\frac{k_0}{2|\varepsilon_{\rm m}|\delta_{\rm skin}}
\int_0^{\infty}\eta(z,t){\rm exp}(-|z|/\delta_{\rm skin})dz \,.
\end{equation}
Figure 1b shows the experimental time-resolved plasmonic
pump-probe interferogram. It depicts the pump-induced changes of
optical properties at the gold-air interface captured by
time-delayed surface plasmon probe pulses and recorded in optical
transmission through the slit. Pump pulses excite the cobalt layer
at zero pump-probe delay time. The most striking feature of a
pump-probe interferogram is the pronounced modulation of plasmonic
interference fringes upon reflection of an acoustic pulse at the
gold-air interface at a pump-probe delay time of $t_0$=39~ps. The
dynamics of the surface dielectric function in Fig.~1c
(reconstructed from the pump-probe interferogram in Fig.~1b, see
Ref.~\cite{SPInterferometry09OE} for details) show that the
acoustic modulation is due to the strong change of the real part
of the dielectric function $\delta\varepsilon^{'}(t)$. The
monotonically increasing background signal in both
$\delta\varepsilon^{'}(t)$ and $\delta\varepsilon^{''}(t)$
describes the slow temperature rise at the gold-air interface on a
time scale of a few tens of picoseconds caused by diffusion of hot
electrons in gold, an effect extensively discussed earlier
\cite{Tas94PRB49}.

\textbf{Reconstruction of ultrashort acoustic strain pulses.}
Information about the dielectric permittivity on an ultrafast time
scale lets us now reconstruct the acoustic strain pulses at the
gold-air interface. Assuming that the shape of the acoustic pulse
does not change during acoustic reflection, i.e. $\eta
(z,t)=\eta(t+z/c_{\rm s})-\eta(t-z/c_{\rm s})$, we obtain a more
convenient expression for $\delta\varepsilon^{'}(t)\simeq
2|\varepsilon_m|^2\delta k_{\rm sp}(t)/k_0$:
\begin{equation}
\label{Epsilon1} \delta\varepsilon^{'}(t)=\frac{|\varepsilon_{\rm
m}|}{\tau_{\rm skin}} \int_{-\infty}^{\infty}\eta(t^{'}){\rm
exp}(-|t-t^{'}|/\tau_{\rm skin}){\rm sgn}(t-t^{'})dt^{'} \,
\end{equation}
with $\eta (t)=\eta (z=0,t-t_0)$ and $\tau_{\rm skin}=\delta_{\rm
skin}/c_{\rm s}$. Taking the time derivative $d\epsilon^{'}/dt$
leads to the following Fredholm integral equation of the second
kind for the strain $\eta(t)$:
\begin{equation}
\label{Strain} \eta(t)=\frac{\tau_{\rm skin}}{2|\varepsilon_{\rm
m}|}\frac{d\varepsilon^{'}}{dt}+\frac{1}{2\tau_{\rm
skin}}\int_{-\infty}^{\infty}\eta(t^{\prime}){\rm
exp}(-|t-t^{\prime}|/\tau_{\rm skin})dt^{\prime} \,.
\end{equation}
Whereas a solution of this integral equation for an arbitrary
pulse is a difficult task, it may be simplified for an ultrashort
acoustic pulse: as long as the duration of the acoustic pulse is
shorter than the acoustic propagation time through the optical
skin depth $\tau_{\rm skin}=3.8$~ps, the first rapidly varying
term in Eq.~(3) dominates and the numerical solution of Fredholm
equation converges after a few iterations.

In Fig.~2a we show the acoustic strain pulse obtained from the
experimental data in Fig.~1c by solving Eq.~(3). The inset in
Fig.~2a illustrates the contributions from the two terms in
Eq.~(3). The first term given by the time-derivative
$d\varepsilon^{\prime}/dt$ determines the overall shape of the
acoustic pulse. The shape is unaffected by small errors in the
determination of the slowly varying background described by the
second integral term.

The retrieved acoustic pulse shape may be well approximated by
simple linear modeling of acoustic propagation, as explained in
detail in Fig.~2b. The initial strain pulse generated in the
cobalt transducer is modeled with a simple exponential ${\rm
exp}(-|t|/\tau_0)$ with $\tau_0=2.5$~ps, followed by a weak
antisymmetric tail reflected from the cobalt-sapphire interface
(acoustic reflectivity R=-0.11). Upon arrival at the gold-air
interface, i.e. after propagation through 120~nm gold, the
acoustic strain pulse suffers from dispersion \cite{Tang11PRB84}.
High-frequency components are left behind the pulse maximum, as
observed in the high frequency ripples in Fig.~2b. Moreover, the
acoustic strain pulse experiences reflection at slightly different
times due to nano-scale surface roughness (SR) at the gold-air
interface. We used atomic force microscopy to quantify SR on all
investigated samples. A typical surface roughness of $\sim$2~nm
(RMS) introduces a distribution of acoustic delay times within
600~fs (FWHM) and smears out the high-frequency ripples of the
acoustic pulse (Fig.~2b).

Excellent quantitative agreement between the experimental data and
the linear model (Fig.~2a) taking into account acoustic dispersion
and surface roughness is achieved by setting the heat penetration
depth in cobalt $\delta_{\rm heat}=c^{{\rm (Co)}}_{\rm s}\tau_0$
to 15~nm with $c^{{\rm(Co)}}_{\rm s}=6.3$~km/s being the the speed
of sound in (0001) direction in hcp-cobalt
\cite{SteinleNeumann99PRB60}. This value is $\sim$1.5 times higher
than the optical skin depth for 400~nm pump pulses in cobalt of
9.6~nm, in agreement with previously reported results for nickel
films \cite{Saito03PRB67}. This similarity of nickel and cobalt is
expected since the electron diffusion is particularly inefficient
in Ni, Co and Fe due to their short electronic mean free paths
\cite{Rendell80PRL45}.

The measured amplitude of the acoustic pulse $\simeq 2\times
10^{-3}$ agrees well with the strain due to thermal expansion of
the laser-heated cobalt layer, estimated from the amount of
absorbed pump energy. Possible contributions due to ultrafast
magnetostriction \cite{KorffSchmising08PRB78} would lead to much
smaller strain values below $1.5\times 10^{-4}$ in cobalt
\cite{Bozorth54PR96}. It is important to note that the acoustic
pulses entering the gold layer preserve their duration and
exponential shape $\sim{\rm exp}(-t/\tau_0)$, but due to the 1.8
times smaller speed of sound in gold they become higher in strain
and spatially contracted. Therefore, particularly high strain
amplitudes in gold should be feasible, in contrast to sapphire,
where acoustic pulses become weaker for the same reason ($c^{{\rm
(sapphire)}}_{\rm s}=11.2$~km/s in (0001) direction).

\textbf{Nonlinear acoustic propagation effects in gold.} In order
to be able to work in the range of relatively high pump fluences,
we had to use (for technical reasons) pump pulses at an optical
wavelength of 800~nm, which were discriminated from probe pulses
at the same wavelength by perpendicular polarization. Using this
arrangement we performed a series of pump-power-dependent
measurements with higher excitation fluence in two samples with an
identical 35~nm thin cobalt transducer, but two different
thicknesses of the gold layer, 120~nm and 220~nm. We clearly
observe nonlinear reshaping of the acoustic pulses in both samples
(Fig.~3). Pulses with larger amplitude are moving faster. The
reshaping is more pronounced in the thicker gold film. Both
observations indicate the importance of nonlinear acoustic
propagation in gold. Our measurements are in excellent
quantitative agreement with the solutions of the non-linear
Korteveg-de Vries (KdV) equation
\begin{equation}
\label{KdV} \frac{\partial\eta}{\partial t}+c_{\rm
s}\frac{\partial\eta}{\partial z}
+\gamma\frac{\partial^3\eta}{\partial z^3}+\frac{C_3}{2\rho
c_s}\eta\frac{\partial\eta}{\partial z}=0 \,.
\end{equation}
These dynamics are governed by the interplay between the acoustic
broadening due to phonon dispersion $\omega(q)=c_{\rm s}q-\gamma
q^3$ with $\gamma=7.41\times 10^{-18}$~m$^3$/s and self-steepening
and reshaping due to the elastic nonlinearity $C_3=-2.63\times
10^{12}$~kg/ms$^2$. See Ref.~\cite{Hao01PRB64} for a definition of
$C_3$ and Ref.~\cite{Behari70JPCSSP3,Hiki66PR144} for linear and
higher order elastic constants in gold; $\rho=19.2$~g/cm$^3$ is
the gold density. Surface roughness is again found to smear out
possible high-frequency acoustic transients on a sub-picosecond
time scale (compare KdV and KdV+SR in Fig.~3). Agreement between
the nonlinear theory and the experiment is simply achieved by
setting a single fit parameter, the initial heat penetration depth
in cobalt $\delta_{\rm heat}=20$~nm (corresponding to
$\tau_0=3.2$~ps initial pulse duration). This value is again 50
percent larger than the optical skin depth in cobalt of 13~nm.

The quantitative agreement between theory and experiment for
120~nm and 220~nm thin gold layers clearly demonstrates that
nonlinear reshaping is dominated by acoustic nonlinearities in
gold. A possible dependence of the initial strain shape in cobalt
on pump power obviously plays a minor role. The amplitude of the
acoustic strain reaches peak values $\eta_{\rm max}=0.01$,
corresponding to a compressional stress $dp=c^2_{\rm
s}\rho\eta_{\rm max}=2.3$~GPa before acoustic reflection and
negative -2.3~GPa tensile stress afterwards. At even higher pump
powers, strain pulses up to 1.5 percent were obtained but
irreversible degradation of the samples was observed, consistent
with elevating the temperature of the cobalt layer close to the
melting point of 1768~K. Control measurements at lower pump
fluences confirmed that the nonlinear reshaping presented in
Fig.~3 is fully reversible.

Intrinsic damping of longitudinal phonons in gold caused by
anharmonic phonon-phonon interactions \cite{Tang11PRB84} becomes
increasingly important for frequencies exceeding 1~THz. However,
in this frequency range and for our experimental geometry the
effect of nano-scale surface roughness entirely masks possible
contributions due to phonon attenuation. Ultrafast acoustic
pump-probe experiments in hybrid multilayer structures with a
thicker gold layer and improved surface quality
\cite{Nagpal09Science325} may be performed to explore the
mechanisms of terahertz phonon attenuation in metals.

\textbf{\textsf{Discussion}}

We have observed nonlinear reshaping of giant ultrashort acoustic
pulses in gold at room temperature by applying femtosecond
time-resolved surface plasmon interferometry. The ability to
generate and quantify high-strain ultrashort acoustic pulses in
hybrid metal/ferromagnet multilayer structures is most relevant
for the emerging field of ultrafast magneto-acoustics \cite
{Scherbakov10PRL105,Thevenard10PRB82,Kim12PRL109}. The most
prominent effects were obtained in the high-strain regime
($\eta_{\rm max}\sim 0.5\%$) using 200~nm nickel films
\cite{Kim12PRL109} in which the nonlinear propagation effects are
likely to be important. We envision using GPa acoustic pulses to
switch magnetization in complex magnetostrictive materials such as
Terfenol-D \cite{KovalenkoMagnetoAcoustics}. This material is
characterized by energy barriers between metastable magnetization
directions in the MPa range and possesses an exceptionally high
magnetostrictive coefficient of $\sim 10^{-3}$, which is about two
orders of magnitude higher than in Ni, Fe or Co
\cite{delaFuente04JPCM16}.

Theoretical estimates by Abrikosov show that our ultrashort GPa
strain pulses can change the distribution function of electrons
caused by phase-matched electron-phonon coupling resulting in the
trapping of a small sub-ensemble of the electrons in the
strain-induced deformation potential $\propto E_{\rm
Fermi}\eta\sim 50$~meV \cite{Abrikosov}. This effect may stimulate
novel applications of time-resolved photoelectron spectroscopy
leading to the observation of transient strongly correlated
electron-phonon states in metals on ultrafast time scales.


\textbf{\textsf{Methods}}

\textbf{Fabrication of gold-cobalt bilayers.} Hybrid
acousto-plasmonic gold/cobalt/sapphire multilayer structures were
manufactured by DC magnetron sputtering (base pressure: $2\times
10^{-7}$~mbar; Ar sputter pressure: $10^{-3}$~mbar) of a (111)
gold layer on top of a hcp-cobalt film deposited on a (0001)
sapphire substrate. Once loaded into the deposition chamber,
sapphire substrates were outgassed for 30 minutes at
$250^{\circ}$C. Deposition of Co and Au layers was carried out
also at $250^{\circ}$C. First, a 35~nm thick Co layer was grown,
then a (111)-textured gold layer a thickness of 120-200~nm was
deposited. Deposition rates were 12~nm/minute for Au and
2~nm/minute for Co. Nanometer surface roughness was characterized
by atomic force microscopy for each investigated sample.

\textbf{Tilted slit-groove microinterferometers.} Plasmonic
slit-groove microinterferometers were milled into the multilayers
with a 30~kV Ga$^+$ focussed ion beam. The slits were 100~nm wide
and extended through both metal layers. The grooves were 200~nm
wide and had a depth of 70 and 120~nm for gold thickness of 120~nm
and 220~nm, respectively. The tilt angle between the slit and the
groove was $10^{\circ}$.

\textbf{Optical measurements.} Time-resolved optical measurements
were performed with an amplified femtosecond  Ti:sapphire laser
(Coherent RegA 250~kHz). The spatial period of the plasmonic
interference pattern slightly varied depending on the angle of
incidence of the probe beam, in agreement with predictions of
geometrical optics. Technical details of the lock-in based
scanning imaging setup used to record plasmonic interferograms
along the slit axes are given elsewhere in
\cite{SPInterferometry09OE}. The Fourier method was used to
extract phase and amplitude of plasmonic interference fringes.


\newpage
\textbf{Acknowledgements}\\
The authors are indebted to V. Shalagatskyi, T. Pezeril, S.
Andrieu, A. Maznev, A. Lomonosov and V.E. Gusev for stimulating
discussions. Financial support by {\it Nouvelle \'{e}quipe,
nouvelle th\'{e}matique de la R\'{e}gion Pays de La Loire}, the
Deutsche Forschungsgemeinschaft (TE770/1 and SFB767)  and by U.S.
DOE Grant No. DE-FG02-00ER15087
and NSF Grants Nos. CHE-0616939 and DMR-0414895 is gratefully acknowledged.\\

\textbf{Author contributions}\\ All authors have contributed to
this paper and agree to its contents.\\

\textbf{Competing interests}\\ The authors declare that they have
no competing financial interests.\\

 \textbf{Correspondence}\\Correspondence and requests for materials should be addressed to
V.T.~(email: vasily.temnov@univ-lemans.fr).\\

\textbf{\textsf{Figure legends}}

\begin{figure}
\includegraphics[width=8.5cm]{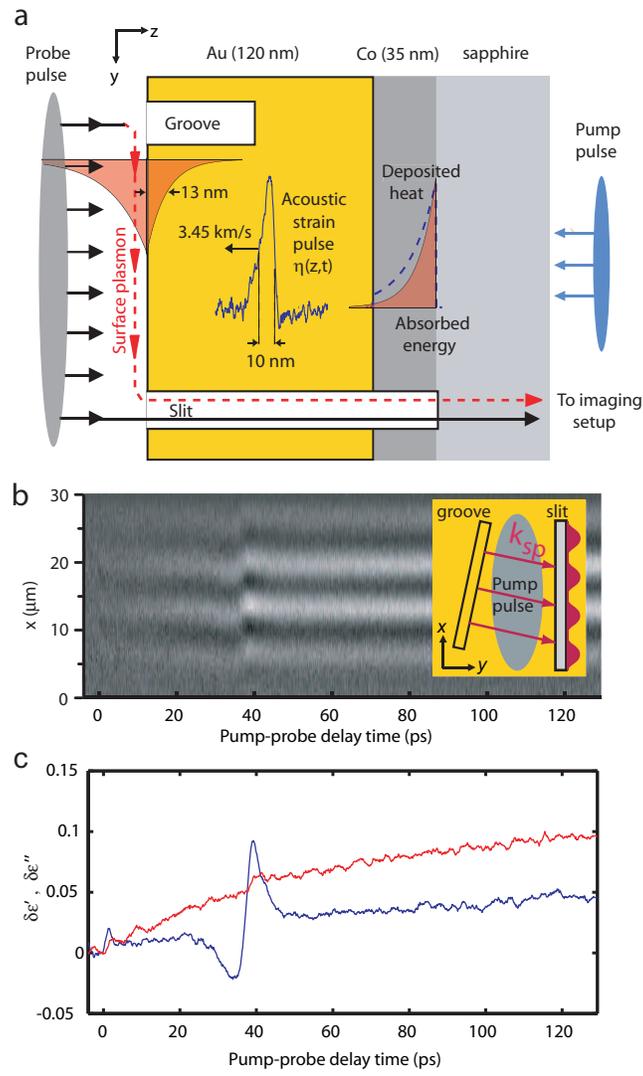} \caption{ {\bf
Acousto-plasmonics in a hybrid gold-cobalt multilayer structure.}
{\bf a,} Schematic drawing of the acousto-plasmonic pump-probe
experiment: surface plasmons propagating at the gold-air interface
probe the reflection of acoustic pulses generated in the
laser-heated cobalt transducer. {\bf b,} Measured
acousto-plasmonic pump-probe interferogram showing a pronounced
shift of the interference fringes upon reflection of an ultrashort
acoustic pulse. The inset illustrates the geometry of the
plasmonic slit-groove interferometer (see
Ref.~\cite{Temnov10NPhoton4,SPInterferometry09OE,Temnov12NPhoton6}
for details). {\bf c,} Ultrafast dynamics of the real (red line)
and imaginary (blue line) parts of the surface dielectric function
extracted from the plasmonic interferogram. Panels {\bf a} and
{\bf b} are adopted from Ref.~\cite{Temnov12NPhoton6}.}
\end{figure}

\begin{figure}
\includegraphics[width=8.5cm]{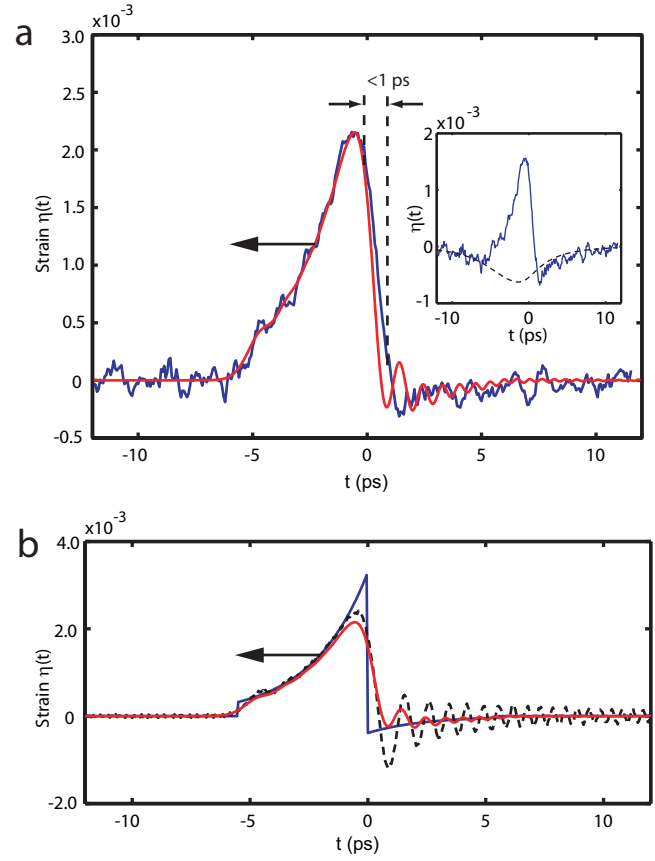} \caption{ {\bf
Acoustic strain pulses at the gold-air interface.} {\bf a,} The
measured strain pulse (blue solid line) launched with a peak
fluence of 7~mJ/cm$^2$ is in excellent agreement with the
theoretical calculation (red solid line). The inset shows the two
terms in Eq.~(3): the first term (blue solid line) and the second
term with opposite sign (dashed line). {\bf b,} Simulation of the
strain pulses generated in cobalt (blue solid line) and after
propagation through a 120~nm gold layer based on a linear model
including dispersion (dashed line), as well as dispersion and
surface roughness (red solid line). The black arrow indicates the
propagation direction of the acoustic pulses (compressional
acoustic strain pulses have positive sign throughout the paper).}
\end{figure}

\begin{figure}
\includegraphics[width=12cm]{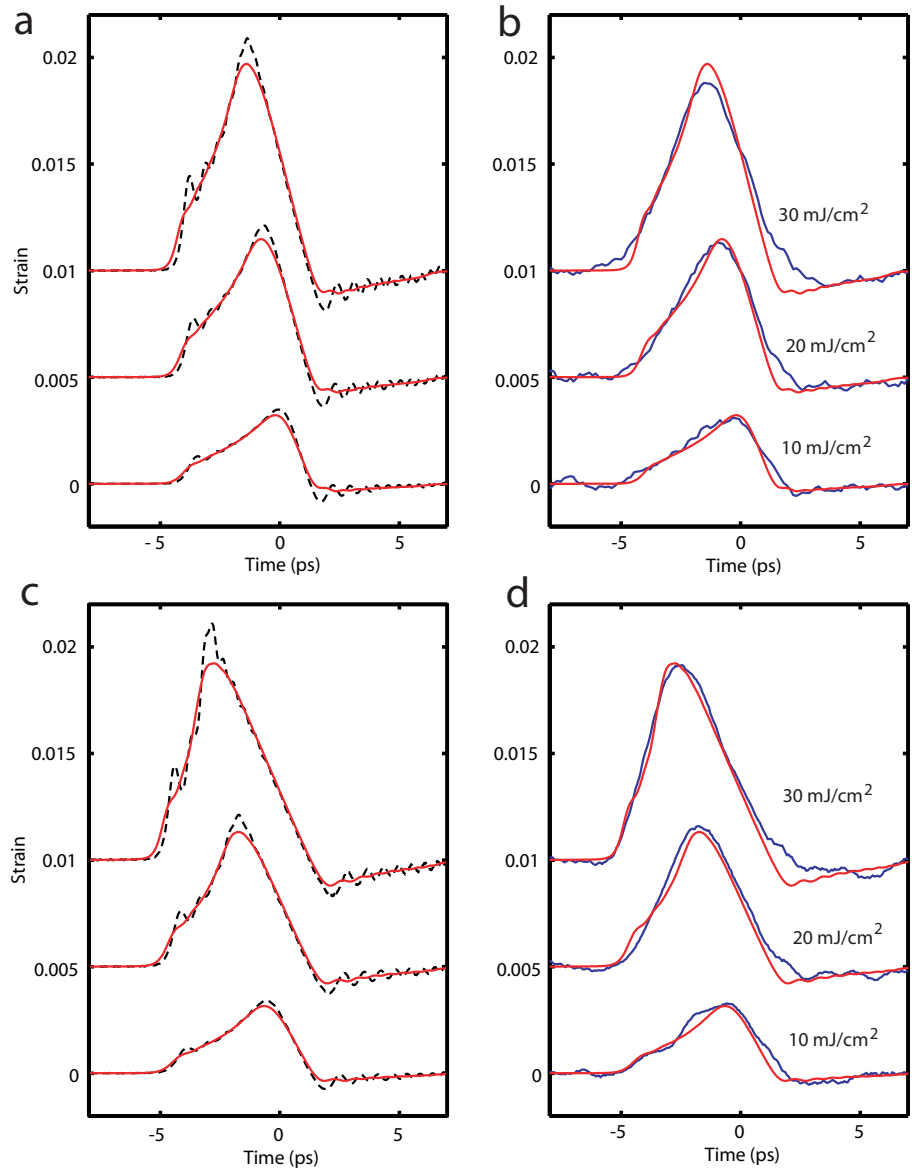} \caption{{\bf Nonlinear
reshaping of giant acoustic strain pulses as a function of
excitation fluence in 120~nm (panels {\bf a} and {\bf b}) and
220~nm (panels {\bf c} and {\bf d}) thick gold films deposited on
a 35~nm thin cobalt transducer.} {\bf a,c} High-frequency
components in the numerical solutions of the Korteveg-de Vries
equation (KdV, dashed line) are smeared out by surface roughness
(KdV+SR, red solid line). {\bf b,d} Experimental curves (blue
solid lines) demonstrate the nonlinear steepening of giant (up to
1 percent) strain pulses, which becomes more pronounced for high
excitation fluences and for thicker gold layers. The quantitative
agreement with the KdV solutions (KdV+SR, red solid lines) is
obtained by adjusting a single fit parameter, namely the 20~nm
heat penetration depth in cobalt at 800~nm pump wavelength. No fit
parameters were used to derive the experimental strain pulses.
Curves for three different values of excitation fluence (10, 20
and 30~mJ/cm$^{2}$) are vertically displaced for clarity. }
\end{figure}


\end{document}